\documentclass[amsmath,amssymb,reprint,aps,prl,showpacs,superscriptaddress,twocolumn,floatfix]{revtex4-1}

\usepackage{graphicx}
\usepackage[dvips]{color}

\begin{document}

\title{Conductance-Driven Change of the Kondo Effect in a Single Cobalt Atom}
\author{D.-J.\ Choi}
\author{M.\ V.\ Rastei}
\affiliation{Institut de Physique et Chimie des Mat\'{e}riaux de Strasbourg$\text{,}$ Universit\'{e} de Strasbourg$\text{,}$ CNRS, 67034 Strasbourg, France}
\author{P.\ Simon}
\affiliation{Laboratoire de Physique des Solides, Universit\'{e} Paris-Sud, CNRS, 91405 Orsay, France}
\author{L.\ Limot}
\email{limot@ipcms.unistra.fr}
\affiliation{Institut de Physique et Chimie des Mat\'{e}riaux de Strasbourg$\text{,}$ Universit\'{e} de Strasbourg$\text{,}$ CNRS, 67034 Strasbourg, France}
\date{\today}

\begin{abstract}
A low-temperature scanning tunneling microscope is employed to build a junction comprising a Co atom bridging a copper-coated tip and a Cu(100) surface. An Abrikosov-Suhl-Kondo resonance is evidenced in the differential conductance and its width is shown to vary exponentially with the ballistic conductance for all tips employed. Using a theoretical description based on the Anderson model, we show that the Kondo effect and the total conductance are related through the atomic relaxations affecting the environment of the Co atom.
\end{abstract}


\maketitle 
The Kondo effect offers the unique opportunity of studying electron correlations in a single object, which are of importance to the emerging field of spintronics where partially filled $d$ or $f$ shells play a central role. This many-body effect arises due to spin-flip processes involving a localized spin carried by an impurity and the electronic spins of the host metal. The Kondo effect has been successfully evidenced through a zero-bias anomaly, also known as the Abrikosov-Suhl-Kondo (ASK) resonance, in the conductance of lithographically defined quantum dots~\cite{goldhaber98,cronenwett98}, carbon nanotubes~\cite{nygard00}, or single molecules~\cite{park02,liang02,scott10} coupled to metallic electrodes.

A fine tuning of the impurity-electrode hybridization could provide an interesting way to test the impact of structural changes on Kondo correlations~\cite{parks07,parks11}. To date, however, the lack of control over this hybridization has resulted in a strongly device-dependent effect. It was recognized quite early \textemdash for example, by exerting an hydrostatic pressure on the host crystal~\cite{schilling73}\textemdash that the Kondo effect is in fact exponentially sensitive to these changes. More recently, by employing a Scanning Tunneling Microscope (STM), similar results were achieved with a single atom. The Kondo effect of the atom was modified through the crystal structure of the host surface~\cite{wahl04,gumbsch10}, or through the strain affecting the atom environment~\cite{quaas04,neel08}. 

Following the seminal point-contact study by Yanson \textit{et al.}~\cite{yanson95}, the ASK resonance has also been observed on rare occasions in well-controlled two-terminal devices where a single atom adsorbed onto a metal surface is brought into contact with the tip of an STM. Surprisingly, the ASK resonance was shown to be little affected by structural changes induced by the tip-atom contact~\cite{neel07,vitali08,bork11}. Motivated by this apparent discrepancy, we revisit in this Letter the tip contact with a Co atom on Cu(100). We show, unlike N{\'e}el \textit{et al.}~\cite{neel07}, that atomic relaxations produced by the tip displacement continuously affect both the ASK resonance and the conductance prior to and after tip contact. A reproducible exponential variation of the ASK line width with the conductance is evidenced and explained through an Anderson-based model. Our findings demonstrate that the Kondo effect of single atoms adsorbed on a metal surface, and more generally their magnetism, may be tuned via the ballistic conductance of the junction.

An STM operating at $4.6$~K in the $10^{-11}$~mbar range was used for the measurements. The Cu(100) was cleaned by sputter/anneal cycles, while single Co atoms were evaporated on the cold surface by heating a high-purity cobalt wire. Etched W tips were cleaned by sputter/anneal cycles and coated with copper \textit{in vacuo} by soft indentations into the clean Cu(100) surface. Tip status was monitored through STM images and controlled tip-atom contacts in order to select tips terminated by a single-copper atom~\cite{limot05}. The differential conductance spectra, $dI/dV(V)$, where $V$ is the sample bias measured with respect to the tip and $I$ is the tunneling current, were acquired via lock-in detection with a bias modulation of amplitude $2$~mV~rms, or less, and a frequency close to $700$~Hz.

\begin{figure}[t]
  \includegraphics[width=0.48\textwidth,clip=]{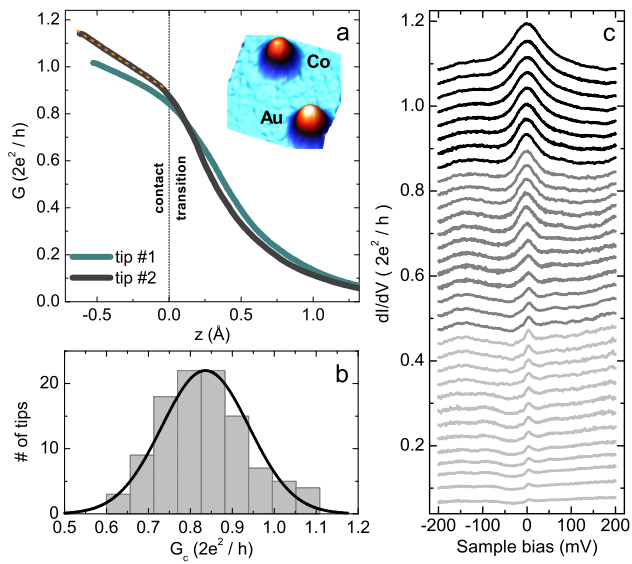}
  \caption{(color online). Evolution of the ASK resonance with conductance. (a) Typical conductance versus tip displacement for two tips above a Co atom on Cu(100) ($V=-200$~mV) and linear fit to $G(z)$ in the contact regime (see text). The vertical dashed line is positioned at $z=0$ where $G=G_c$. Inset: Constant-current image of Co and Au atoms on Cu(100) ($13\times13$~{\AA}$^2$, $-200$~mV, $1$~nA). Apparent heights are $1.1$ and $1.2$~{\AA} for Co and Au, respectively. The contact conductances amount to $0.84$ and $0.72$, respectively. (b) Histogram of $G_c$ for roughly $100$ tips. (c) $dI/dV$ spectra from tunnel (solid light-grey lines) to contact (solid black lines), including spectra in the transition regime affected by atomic relaxations (solid dark-grey lines). All the data is fully reversible with tip displacement.}
\label{fig1}
\end{figure}

\begin{figure}[t]
  \includegraphics[width=0.48\textwidth,clip=]{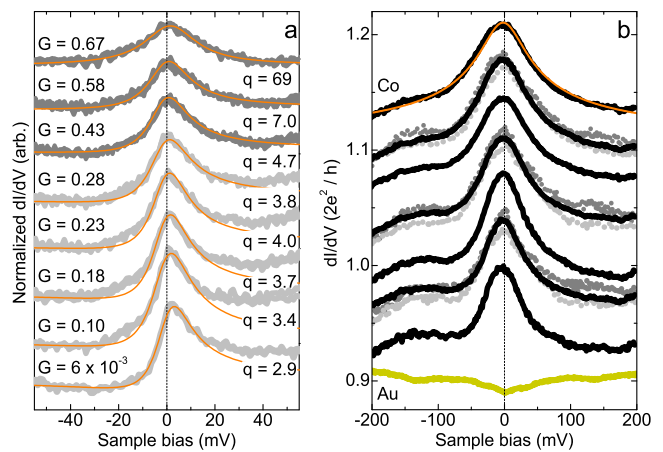}
  \caption{(color online). ASK resonance up close. (a) $dI/dV$ spectra near the tunnel-to-transition regime. Spectra are renormalized by the conductance where the feedback loop was open (indicated on the figure). The solid lines are Fano fits as described in the text and $q$ is the corresponding Fano parameter of each single fit. The ASK resonance is centered close to $\epsilon_{\text{K}}=-1$~meV. (b) $dI/dV$ spectra in the contact regime. The solid orange line corresponds to the Frota function described in the text. We have included spectra acquired with two other tips (gray dotted lines, vertically displaced upward by $0.005$; light-gray dotted lines displaced downward by $-0.005$). The spectrum acquired in contact with a non-magnetic Au atom on Cu(100) is featureless. The vertical dashed line is positioned at zero bias.}
\label{fig2}
\end{figure}

Figure~\ref{fig1}(a) presents the evolution of the conductance ($G=I/V$) for two given copper-coated tips as they are vertically displaced towards the center of a Co atom on Cu(100) [inset of Fig.~\ref{fig1}(a)]. A tip displacement (noted $z$ hereafter) of $z=0$ defines the boundary between the transition regime ($z>0$) and the contact regime ($z<0$). In the transition regime, electrons tunnel between the tip and the substrate, but the conductance is influenced by atomicscale relaxations~\cite{limot05,vitali08,ternes11}; the structural relaxations vanish once $z\gtrsim 1$~{\AA} [not shown in Fig.~\ref{fig1}(a)]. The $z=0$ boundary is deduced through the contact conductance ($G_{\text{c}}$), which is determined following a geometrical approach described elsewhere~\cite{neel09}. An average of $\langle G_c\rangle=0.84\pm0.09$ (in units of $2e^2/h$) is found for Co/Cu(100) from the statistical survey presented in Fig.~\ref{fig1}(b). The contact formation between the tip and the Co atom is therefore highly reproducible and reversible. This follows from the larger stiffness of the atom bond to the surface compared to the pristine surface~\cite{limot05}. In Fig.~\ref{fig1}(c), we present the evolution of the Kondo effect of Co/Cu(100) in the presence of a copper-coated tip. To do so, we freeze the geometry of the junction by opening the feedback loop at selected conductances and acquire a $dI/dV$ spectrum. These conductances correspond in Fig.~\ref{fig1}(c) to the value of the $dI/dV$ at $-200$~mV. Several orders of magnitude for the conductance were covered in this way, corresponding to a tip excursion of about $5.5$~{\AA}. In Fig.~\ref{fig1}(c), we focus on the final $2$~{\AA} where substantial spectral changes are evidenced. An ASK resonance is detected at the Fermi level, but compared to the previous study~\cite{neel07}, the line width evolves continuously with increasing conductance. The color code employed in Fig.~\ref{fig1}(c) reflects the different spectral regimes described in detail hereafter. 

The resonance in the transition and tunneling regimes ($z<0$) is shown in Fig.~\ref{fig2}(a), and may be described by a Fano line $dI/dV\propto(q+\epsilon)^2/(1+\epsilon^2)$~\cite{ujsaghy00,plihal01}, where $\epsilon=(eV-\epsilon_{\text{K}})/k_{\text{B}}T_{\text{K}}$; $q$ and $\epsilon_{\text{K}}$ are the asymmetry parameter and the energy position of the resonance, respectively. The full width at half maximum (FWHM) of the ASK resonance is therefore equivalent to $2k_{\text{B}}T_{\text{K}}$. The resonance is detected as a steplike feature with a Fano parameter ranging from $q=2.9$ to $4.7$ up to $G=0.43$ [light-grey spectra in Fig.~\ref{fig2}(a)], which corresponds to $z=0.5$~{\AA}. Starting here, the Fano parameter increases rapidly until at $G=0.67$ ($z=0.25$~{\AA}) the ASK line shape becomes peaklike [dark-grey spectra in Fig.~\ref{fig2}(a)]. Interestingly, the Kondo effect of Co has been studied on various copper surfaces by tunneling spectroscopy and shown to be diplike on Cu(111) (corresponding to $q=0$)~\cite{knorr02,vitali08}, steplike on Cu(100) ($q\sim1$)~\cite{knorr02}, or peaklike on Cu(110) ($q\gg 1$)~\cite{gumbsch10}. The spectral line shape is therefore sensitive to modifications of the surface electronic structure~\cite{plihal01}, which imply also changes for the impurity $d$-band structure. We then assign the changes of $q$ in Fig.~\ref{fig2}(a) to a modification of the Co and Cu(100) electronic structure induced by the tip proximity. A microscopic description of the changes observed for $q$ would require an atomistic approach~\cite{lucignano09,jacob09} beyond the scope of the present study.

As shown in Fig.~\ref{fig2}(b), the spectral line shape remains peaklike in the contact regime up to the highest conductances explored. A very satisfying agreement is found between our $dI/dV$ data and simulations of the Kondo line shape [solid line in Fig.~\ref{fig2}(b)] carried out with a Frota function~\cite{frota92}, which closely reproduces the exact solution that can be obtained with numerical-renormalization group theory~\cite{pruser11}. Along with Co, we also monitored the $dI/dV$ spectra on non-magnetic Au and Cu atoms, where, as expected, no ASK resonance is present [Fig.~\ref{fig2}(b)].

\begin{figure}
  \includegraphics[width=0.48\textwidth,clip=]{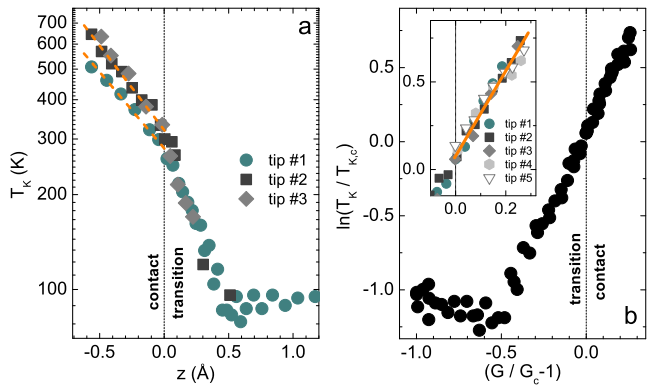}
  \caption{(color online). Evolution of the Kondo temperature. (a) $T_{\text{K}}$ versus tip displacement. The solid lines correspond to the fits described in the text. The Kondo temperature was extracted by Fano fits [see Fig.~\ref{fig2}(a)], or, alternatively, by measuring in the contact regime the FWHM of the ASK resonance. (b) $T_{\text{K}}$ versus conductance for various tips. The vertical dashed line is positioned at $G=G_c$. $T_{\text{K}}$ is the Kondo temperature at $z=0$ and experimentally determined to be close to $300$~K for all tips employed. Inset: $T_{\text{K}}$ in the contact regime for five tips labeled $1,2,3,4,5$ which have, respectively, a contact conductance of $G_c=0.87,0.90,0.91,0.90,0.85$. The solid line in the contact regime corresponds to the fit described in the text.}
\label{fig3}
\end{figure}

The changes in the line shape are also characterized by a concomitant increase of the FWHM, hence of $T_{\text{K}}$, which we plot in Fig.~\ref{fig3}(a) as a function of tip displacement. For all tip apices employed, in the tunneling regime we find $T_{\text{K}}=90$~K in agreement with previous studies~\cite{knorr02,neel07}. Starting from $z=0.5$~{\AA} however, $T_{\text{K}}$ increases and rapidly reaches a temperature close to $300$~K at $z=0$. In the contact regime ($z<0$), $T_{\text{K}}$ increases exponentially with $z$, but is tip-dependent. For some tips, values up to $650$~K are found at $z=-0.6$~{\AA} (higher tip excursions resulted in tip instabilities). The maximum value found for $T_{\text{K}}$ is close to the one for Co in bulk copper~\cite{daybell68,pruser11}, despite the lower coordination of Co in our junction~\cite{note1}. Quite remarkably, the exponential dependency of the Kondo temperature on $G$ is instead reproducible as shown in Fig.~\ref{fig3}(b) for a collection of tips. To further exemplify this, Fig.~\ref{fig2}(b) shows ASK resonances acquired with other tips (solid light gray and gray lines). The exponential dependency holds in the transition regime and eventually breaks down below $G=0.4$ ($z=0.5$~{\AA}). When moving to even lower conductances, $T_{\text{K}}$ slightly increases again. 

In view of our findings, heating effects play a minor role in the changes observed for the resonance line width. Firstly, the line width starts increasing in the transition regime where heating effects are negligible~\cite{sperl10}. Secondly, if heating were significant in the contact regime, the line shape of a $dI/dV$ spectrum would profoundly differ from the predicted spectral function [$i.\ e.$ Frota simulation in Fig.~\ref{fig2}(b)] as heating is a bias-dependent effect. If we recall the typical length scales involved in the Kondo screening ($\xi_{\text{K}}$) and in the inelastic scattering, it is not surprising that $T\ll T_\text{K}$. The spatial range of the Kondo effect for Co/Cu(100) is in fact $\xi_{\text{K}}=\hbar v_F/k_BT_{\text{K}}\simeq30$~nm ($v_F$: Fermi velocity of copper)~\cite{sorensen96}. At this reduced scale, inelastic electron-electron scattering is responsible for the temperature of the electronic bath~\cite{anderson79,roukes85,ditusa92}, the scattering length amounting then to $\xi_{e-e}=v_F \tau_{e-e}$, where $\tau_{e-e}^{-1}$ is the energy-dependent scattering rate~\cite{campillo99}. For the biases employed, $\xi_{e-e}$ is equivalent to at least $900$~nm, in other words the inelastic scattering of electrons is negligible as $\xi_{\text{K}}\ll\xi_{e-e}$.

To understand how $T_{\text{K}}$ relates to $G$, we treat the Kondo effect of a single Co impurity within an effective spin-$1/2$ Anderson impurity model without orbital degeneracy~\cite{ujsaghy00}, where the Kondo temperature is ~\cite{schrieffer66,daybell68}
\begin{equation}
k_{\text{B}}T_{\text{K}} \simeq \sqrt{\Gamma U/2} \exp\{-\frac{\pi(\epsilon+U)  |\epsilon|}{2\Gamma U} \}.
\label{eq1}
\end{equation}
This equation relates the Kondo temperature $T_{\text{K}}$ to the on-site Coulomb repulsion $U$, the half width of the hybridized $d$-level $\Gamma$ and its energy $\epsilon$. Density functional theory (DFT) is used to map the ``real'' system [e.g. Co/Cu(100)] onto the spin-$1/2$ effective Anderson model~\cite{ujsaghy00}. $\epsilon$ and $U$ are determined through the band structure of Co by estimating the $d$-level occupation and the exchange splitting between majority and minority $d$-bands. Based on existing DFT~\cite{huang06,neel07,vitali08}, these bands shift in energy when strain is exerted onto Co through a tip displacement.  Since $\Gamma$ is nearly constant (see below), the main contributions to an increased Kondo temperature in Eq.~(\ref{eq2}) come from changes in $\epsilon$ and $U$ that are driven by the forces acting between the Co atom and the tip-apex atom. The changes we detect in $T_{\text{K}}$, and presumably also in the Fano parameter $q$, indicate that adhesive forces set in at least $0.5$~{\AA} prior to contact in agreement with previous studies~\cite{limot05,ternes11}. 

\begin{figure}
  \includegraphics[width=0.33\textwidth,clip=]{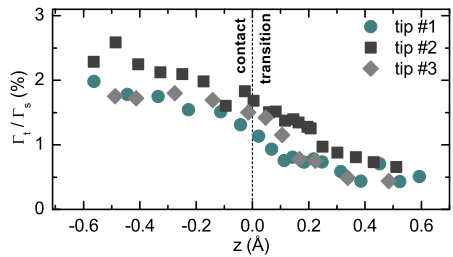}
  \caption{(color online). Relative coupling $\Gamma_t/\Gamma_s$ for various tips. The vertical dashed line is centered at $z=0$.}
\label{fig4}
\end{figure}

To quantify the impact of tip displacement on $\Gamma$, we have determined in Fig.~\ref{fig4} the relative $s-d$ coupling of the atom to the surface ($\Gamma_s$) and tip ($\Gamma_t$) based on the magnitude of the differential conductance at the Fermi level. In the limit of $T\ll T_{\text{K}}$~\cite{glazman88,ng88,pustilnik01}, which is fulfilled here, we have
\begin{equation}
dI/dV(V=0) \approx (4\Gamma_t\Gamma_s)/(\Gamma_s+\Gamma_t)^2+(dI/dV)_{\text{el}}
\label{eq2}
\end{equation}
(in units of $2e^2/h$). The first term accounts for the ASK resonance, while the second term corresponds to a background contribution and amounts to at least $90\%$ of the total conductance [Fig.~\ref{fig1}(c)]. A majority of electrons have therefore a ballistic conductance with a transmission most likely governed by the $4s$ state of Cu and the $3d$ and $4s$ states of Co. Only a minor fraction of the overall electrons is instead affected by Kondo correlations ($G_c$ is in fact little dependent on bias). This is favored in our setup by the small $\Gamma_t/\Gamma_s$ ratio of $3\%$ or lower reported in Fig.~\ref{fig4}. This ratio smoothly increases from tunnel to contact, but only the weaker coupling $\Gamma_t$ changes significantly so that the overall $\Gamma$ is practically unaffected by the tip displacement. The total coupling is constant and $\Gamma=\Gamma_s+\Gamma_t\simeq\Gamma_s$. This setup recalls that of a mechanically-controlled break-junction device with asymmetric lead couplings~\cite{parks07}.

To fully link Eq.~(\ref{eq1}) to our experimental observations, we need first to formally express the changes in $T_{\text{K}}$ with the $d$-band structure of Co, hence with $z$. We introduce the occupation of the $d$-level $n$, which in the spin-$1/2$ effective Anderson model can vary between $0$ (empty orbital) and $2$ (double occupancy). The Kondo regime corresponds to $0.8<n< 1.2$~\cite{zlatic85} with approximately one unpaired electron in the $d$ level. In the tunneling regime, $n$ is estimated between $n=1.16$~\cite{wahl04} and $n=0.91$~\cite{gumbsch10} for Co on Cu(100). In the contact regime, the $d$-level occupation is unchanged~\cite{neel07} or increases slightly due to a downward shift of the minority $d$-band~\cite{huang06}, but cannot increase to higher values than $1.2$ in order to preserve the Kondo physics. Following then {\'U}js{\'a}ghy \textit{et al.} \cite{ujsaghy00}, we relate $\epsilon$ and $U$ by $n=-\epsilon/U+1/2$ and remark that the range of possible values for the $\epsilon/U$ ratio is therefore strongly reduced. In the following, we suppose then that the $\epsilon/U$ ratio is constant~\cite{note2}. This translates into a simplified expression for Eq.~(\ref{eq1})
\begin{equation}
\ln(T_{\text{K}}/T_{\text{K,c}}) \simeq (\epsilon-\epsilon_c)/\tilde{\Gamma}=(-\beta\kappa/\tilde{\Gamma})z,
\label{eq3}
\end{equation}
where $\tilde{\Gamma}=2\Gamma/\pi(1.5-n)$ and $T_{\text{K,c}}$ is the Kondo temperature at $z=0$~\cite{note3}. Equation~(\ref{eq3}) explicitly shows that the $d$-band displacement relative to the equilibrium position at $z=0$ (noted $\epsilon_c$) exponentially changes the Kondo temperature in the contact regime. Remarking that this band shift is driven by changes in the Co-Cu distances due to the tip displacement, we have expressed $\epsilon(z)-\epsilon_c$ within a tight binding approximation. The transfer integral is approximated by $-\beta\kappa z$~\cite{rastei07}, $\beta$ being proportional to the cobalt $d$ bandwidth and $\kappa$ a structure dependent decay constant; in transition metals, $\kappa$ is related to the bulk modulus and is $1$~{\AA}$^{-1}$ or higher~\cite{desjonqueres93}. As $\beta$ and $\tilde{\Gamma}$ relate to properties of the Co atom, the tip dependency evidenced in Fig.~\ref{fig3}(a) is carried by $\kappa$. On average we find a slope of $\beta\kappa/\tilde{\Gamma}=-1.2\pm0.1$~{\AA}$^{-1}$. From the band shift $\epsilon-\epsilon_c$ predicted by DFT in the contact regime we estimate $\beta\kappa\approx 0.5$~eV{\AA}$^{-1}$~\cite{vitali08}, which, based on our findings, would then yield a reasonable $s-d$ coupling of $\Gamma\approx0.3$~eV for a $d$-level occupation of $n\approx 1$. 

We may now link $T_{\text{K}}$ to the conductance in the contact regime. As mentioned above [see discussion based on Eq.~(\ref{eq2})], the conductance is mainly governed by ballistic electrons. We then express the dependence of $G$ on $z$ within a linear expansion $G/G_c=1+\kappa_0 z$, where the decay constant is $\kappa_0=(dG/dz)_{z=0}/G_c$. This decay constant is extracted from $G(z)$ [dashed lines in Fig.~\ref{fig1}(a)] and is tip-dependent. The average value for the tips employed amounts to $\kappa_0=0.45$~{\AA}$^{-1}$ so that $\kappa_0 z\ll1$ for the tip displacements considered here. By substituting this expression for $G$ in Eq.~(\ref{eq3}) we arrive then to
\begin{equation}
\ln(T_{\text{K}}/T_{\text{K,c}})=(\beta/\tilde{\Gamma})(\kappa/\kappa_0)(G/G_c-1).
\label{eq4}
\end{equation}
We have used Eq.~(\ref{eq4}) to fit the data of Fig.~\ref{fig3}(b) (solid line) and find that $(\beta/\tilde{\Gamma})(\kappa/\kappa_0)=2.4\pm0.1$ for all tips employed. From the data, we extract $\kappa_0\approx0.5$~{\AA}$^{-1}$, which is consistent with the previous estimate. Although $\kappa$ and $\kappa_0$ are tip dependent, the changes of $T_{\text{K}}$ with $G$ are not, which indicates that the ratio between the two decay constants must cancel out the tip dependency. More generally, our findings show that in this regime the $d$-band of Co/Cu(100), and therefore $T_{\text{K}}$, can be exponentially tuned through the conductance of the junction, regardless of tip structure. This follows from the highly reproducible contact geometry, and hence reproducible $G_c$, achieved with this experimental setup. The above relation also applies in the transition regime, but only over a finite conductance range as $\kappa_0$ increases due to the presence of the tunneling barrier. We find in fact $\kappa_0=2.15$~{\AA}$^{-1}$ in the tunneling regime where $G\propto\exp(-\kappa_0 z)$, which then limits the validity of Eq.~(\ref{eq4}) to $G\gtrsim 0.6$. For a full description of the transition regime, an analytical expression of $G(z)$ would be needed.

In summary, by engineering a well-defined quantum point-contact with an STM, we have shown that the Kondo effect of a single Co atom may be exponentially tuned through the otal conductance of the junction. This follows from the atomic relaxations present in the junction, which provide a link between conductance and Kondo physics. Our findings should apply to other single-impurity Kondo systems and, more generally, open the interesting perspective of controlling, and even activating the magnetism of single impurities on surfaces through the conductance.

\acknowledgments{
We thank J.\ Hieulle for his help during preliminary measurements. We thank R.\ Jalabert,  W.\ A.\ Hofer and V.\ S.\ Stepanyuk for fruitful discussions, and particularly P.\ A.\ Ignatiev for sharing unpublished results.}


\begin{thebibliography}{nnnyys}

\bibitem{goldhaber98}  D.\ Goldhaber-Gordon \textit{et al.} , Nature \textbf{391}, 156 (1998).

\bibitem{cronenwett98}  S.\ M.\ Cronenwett, T.\ H.\ Oosterkamp, and L.\ P.\ Kouwenhoven, Science \textbf{281}, 540 (1998).

\bibitem{nygard00} J.\ Nyg{\aa}rd \textit{et al.}, Nature \textbf{408}, 342 (2000).

\bibitem{park02} J.\ Park \textit{et al.}, Nature \textbf{417}, 722 (2002).

\bibitem{liang02} W.\ Liang \textit{et al.}, Nature \textbf{417}, 725 (2002).

\bibitem{scott10}  G.\ D.\ Scott and D.\ Natelson, ACS Nano \textbf{4}, 3560 (2010).

\bibitem{parks07}  J. J. Parks \textit{et al.}, Phys. Rev. Lett. \textbf{99}, 026601 (2007). 

\bibitem{parks11}  J.\ J.\ Parks \textit{et al.}, Science \textbf{328}, 1370 (2010).

\bibitem{schilling73}  J.\ S.\ Schilling and W.\ B.\ Holzapfel, Phys. Rev. B \textbf{8}, 1216 (1973).

\bibitem{wahl04}  P.\ Wahl \textit{et al.}, Phys. Rev. Lett. \textbf{93}, 176603 (2004). 

\bibitem{gumbsch10}  A.\ Gumbsch \textit{et al.}, Phys. Rev. B \textbf{81}, 165420 (2010).

\bibitem{quaas04} N.\ Quaas, M.\ Wenderoth, A.\ Weismann, R.\ G.\ Ulbrich, and K.\ Sch{\"o}nhammer, Phys. Rev. B \textbf{69}, 201103 (2004).

\bibitem{neel08}  N.\ N{\'e}el \textit{et al.}, Phys. Rev. Lett. \textbf{101}, 266803 (2008).

\bibitem{yanson95}  I.\ K.\ Yanson \textit{et al.}, Phys. Rev. Lett. \textbf{74}, 302 (1995).

\bibitem{neel07}  N.\ N{\'e}el, J.\ Kr{\"o}ger, L.\ Limot, K.\ Palotas, W.\ A.\ Hofer, and R.\ Berndt, Phys. Rev. Lett. \textbf{98}, 016801 (2007). 

\bibitem{vitali08}  L.\ Vitali \textit{et al.}, Phys. Rev. Lett. \textbf{101}, 216802 (2008). 

\bibitem{bork11}  J.\ Bork \textit{et al.}, Nat. Phys. \textbf{7}, 901 (2011). 

\bibitem{limot05}  L.\ Limot, J.\ Kr{\"o}ger, R.\ Berndt, A.\ Garcia-Lekue, and W.\ A.\ Hofer, Phys. Rev. Lett. \textbf{94}, 126102 (2005). 

\bibitem{ternes11}  M.\ Ternes \textit{et al.}, Phys. Rev. Lett. \textbf{106}, 016802 (2011). 

\bibitem{neel09}  N.\ N{\'e}el, J.\ Kr{\"o}ger, and R.\ Berndt, Phys. Rev. Lett. \textbf{102}, 086805 (2009). 

\bibitem{ujsaghy00}  O.\ {\'U}js{\'a}ghy, J.\ Kroha, L.\ Szunyogh, and A.\ Zawadowski, Phys. Rev. Lett. \textbf{85}, 2557 (2000). 

\bibitem{plihal01}  M.\ Plihal and J.\ W.\ Gadzuk, Phys. Rev. B \textbf{63}, 085404 (2001). 

\bibitem{knorr02}  N.\ Knorr, M.\ A.\ Schneider, L.\ Diekh{\"o}ner, P.\ Wahl, and K.\ Kern, Phys. Rev. Lett. \textbf{88}, 096804 (2002).  

\bibitem{lucignano09}  P.\ Lucignano, R.\ Mazzarello, A.\ Smogunov, M.\ Fabrizio, and E.\ Tosatti, Nat. Mater. \textbf{8}, 563 (2009). 

\bibitem{jacob09} D.\ Jacob, K.\ Haule, and G.\ Kotliar, Phys. Rev. Lett. \textbf{103}, 016803 (2009).

\bibitem{frota92} H.\ O.\ Frota, Phys. Rev. B \textbf{45}, 1096 (1992).

\bibitem{pruser11}  H.\ Pr{\"u}ser \textit{et al.}, Nat. Phys. \textbf{7}, 203 (2011). 

\bibitem{sperl10}  A.\ Sperl, J.\ Kr\"oger, and R.\ Berndt, Phys. Rev. B, \textbf{81}, 035406 (2010).

\bibitem{sorensen96}  E.\ S.\ S\o{}rensen and I.\ Affleck, Phys. Rev. B \textbf{53}, 9153 (1996).

\bibitem{anderson79}  P.\ W.\ Anderson, E.\ Abrahams, and T.\ V.\ Ramakrishnan, Phys. Rev. Lett. \textbf{43}, 718 (1979).

\bibitem{roukes85} M.\ L.\ Roukes, M.\ R.\ Freeman, R.\ S.\ Germain, R.\ C.\ Richardson, and M.\ B.\ Ketchen, Phys. Rev. Lett. \textbf{55}, 422 (1985).

\bibitem{ditusa92} J.\ F.\ DiTusa, K.\ Lin, M.\ Park, M.\ S.\ Isaacson, and J.\ M.\ Parpia, Phys. Rev. Lett. \textbf{68}, 1156 (1992).

\bibitem{campillo99} I.\ Campillo, J.\ M.\ Pitarke, A.\ Rubio, E.\ Zarate, and P.\ M.\ Echenique, Phys. Rev. Lett. \textbf{83}, 2230 (1999).

\bibitem{daybell68} M.\ D.\ Daybell and W.\ A.\ Steyert, Rev. Mod. Phys. \textbf{40}, 380 (1968). 

\bibitem{note1} Estimates of $T_{\text{K}}$ change depending on the analytical function used to reproduce the ASK resonance. For example, if one follows \cite{scott10}, our $T_{\text{K}}$ estimates need to be divided by $\sqrt{2}$; if one follows~\cite{pruser11} [see Frota fit in Fig.~\ref{fig2}(b)], $T_{\text{K}}$ estimates need to be divided by a factor $2$. The evolution of $T_{\text{K}}/T_{\text{K,c}}$ with conductance is robust to the method used to estimate the Kondo temperature.

\bibitem{schrieffer66}  J.\ R.\ Schrieffer and P.\ A.\ Wolff, Phys. Rev. \textbf{149}, 491 (1966). 

\bibitem{huang06}  R.\ Z.\ Huang \textit{et al.}, Phys. Rev. B, \textbf{73}, 153404 (2006).

\bibitem{glazman88}  L.\ I.\ Glazman and M.\ E.\ Raikh, JETP Lett. \textbf{47}, 452 (1988).

\bibitem{ng88}  T.\ K.\ Ng and P.\ A.\ Lee, Phys. Rev. Lett. \textbf{61}, 1768 (1988). 

\bibitem{pustilnik01}  M.\ Pustilnik and L.\ I.\ Glazman, Phys. Rev. Lett. \textbf{87}, 216601 (2001). 

\bibitem{zlatic85}  V.\ Zlati{\'c}, B.\ Horvati{\'c}, and D.\ {\u{S}}ok{\u{c}}evi{\'c}, Z. Phys. B \textbf{59}, 151 (1985).

\bibitem{note2}  The continued evolution of $U$ in the contact regime should benefit from further study.

\bibitem{note3} We neglect changes in $U$ in the prefactor of Eq.~(\ref{eq1}), which would yield a first order correction of $\tilde{\Gamma}/2\epsilon_c\ll 1$ to Eq.~(\ref{eq3}).

\bibitem{rastei07}  M.\ V.\ Rastei \textit{et al.}, Phys. Rev. Lett. \textbf{99}, 246102 (2007). 

\bibitem{desjonqueres93}  M.-C.\ Desjonqueres and D.\ Spanjaard, \textit{Concepts in Surface Physics} (Springer-Verlag, Berlin, 1993).

\end{thebibliography}
\end{document}